\documentclass[useAMS,usenatbib]{mn2e}

\usepackage{graphicx}

\title[VLBI detection of MAXI~J1836--194]{VLBI detection of the Galactic black hole binary candidate MAXI J1836--194 }
\author[Yang et al.]{Jun\,Yang$^{1}\thanks{E-mail:yang@jive.nl}$,
Yonghua\,Xu$^{2,3}$,
Zhixuan\,Li$^{2,3}$,
Zsolt\,Paragi$^{1}$,
Robert\,M.\,Campbell$^{1}$, \and
Leonid\,I.\,Gurvits$^{1,4}$,
Zhiqiang\,Shen$^{5,6}$,
Xiaoyu\,Hong$^{5,6}$,
Bo\,Xia$^{5,6}$, Fengchun\,Shu$^{5,6}$
\\
\\
$^{1}$Joint Institute for VLBI in Europe, Postbus 2, 7990 AA Dwingeloo, The~Netherlands \\
$^{2}$Yunnan Astronomical Observatory, NAOC, 650011 Kunming, P.R.~China\\
$^{3}$Key Laboratory for the Structure and Evolution of Celestial Objects, CAS, P.R.~China \\
$^{4}$Department of Astrodynamics and Space Missions, Delft University of Technology, 2629~HS Delft, The~Netherlands \\
$^{5}$Shanghai Astronomical Observatory, CAS, 200030 Shanghai, P.R.~China \\
$^{6}$Key Laboratory of Radio Astronomy, CAS, P.R.~China \\
}

\begin{document}

\date{Accepted 2012 June xx. Received 2012 June xx; in original form 2012 June xx}

\pagerange{\pageref{firstpage}--\pageref{lastpage}} \pubyear{2012}
\maketitle
\label{firstpage}

\begin{abstract}
The X-ray transient MAXI~J1836--194 is a newly-identified Galactic black hole binary candidate. As most X-ray transients, it was discovered at the beginning of an X-ray outburst. After the initial canonical X-ray hard state, the outburst evolved into a hard intermediate state and then went back to the hard state. The existing RATAN-600 radio monitoring observations revealed that it was variable on a timescale of days and had a flat or inverted spectrum, consistent with optically thick synchrotron emission, possibly from a self-absorbed jet in the vicinity of the central compact object. We observed the transient in the hard state near the end of the X-ray outburst with the European VLBI Network (EVN) at 5 GHz and the Chinese VLBI Network (CVN) at 2.3 and 8.3 GHz. The 8.3 GHz observations were carried out at a recording rate of 2048~Mbps using the newly-developed Chinese VLBI data acquisition system (CDAS), twice higher than the recording rate used in the other observations. We successfully detected the low-declination source with a high confidence level in both observations. The source was unresolved ($\leq$0.5~mas), which is in agreement with an AU-scale compact jet.
\end{abstract}


\section{Introduction}
\label{sec1}

MAXI~J1836--194 is a new X-ray transient, discovered on 2011 August 30 \citep{neg11}. The transient was classified as a black hole candidate, because a relativistically broadened iron emission line was revealed by \emph{Suzaku} observations \citep{rei12} during its first known X-ray outburst. The X-ray outburst started from the canonical low/hard X-ray state, then it evolved to the hard intermediate state and finally went back to the low/hard state. The outburst was classified by \citet{fer12} as a ``failed" outburst because it never entered high/soft X-ray state.

The transient source MAXI~J1836--194 was quite bright in radio and infrared. The Karl~G.~Jansky Very Large Array (VLA) and the RATAN-600 radio observations revealed an optically-thick spectrum with a flux density of 20~--~50 mJy between 5 and 8~GHz \citep{mil11, tru11}. The RATAN-600 5-GHz light curve indicates that its emission varied frequently by a factor of two on the timescale of days. Very Large Telescope (VLT) observations \citep{rus11} revealed a secular mid-infrared brightening, reaching 57$\pm$1~mJy at 12.0~$\mu$m on 2011 October 11, making it one of the brightest black hole candidate ever detected at 10 microns.

In view of the rare occurrence of such a bright transient source associated with a black hole candidate, we conducted milliarcsecond-resolution VLBI (Very Long Baseline Interferometry) observations of the radio counterpart of MAXI~J1836--194 just after the mid-infrared brightening. In this paper, we report the first VLBI detection of its compact radio core.

\begin{table*}
\caption{Summary of the VLBI observations.}
\label{tab1}
\scriptsize
\setlength{\tabcolsep}{6pt}

\centering
\begin{tabular}{ccccccccc}
\hline
Proj. Code & Date      & UT Range       & Stations          & Baseline Lengths
                                                                        & Freq.
                                                                                & Duration
                                                                                       & Data Rate       \\
         & (yyyy-mm-dd) &(hh:mm -- hh:mm)&                & (km)         & (GHz) &  (h) & (Mbps)  \\
\hline
CHIN06A  & 2011-10-10   & 08:30 -- 13:00 &     KmShUr     &  1920\,--\,3249      & 8.3   &  4   & 2048    \\
CHIN06B  & 2011-10-10   & 08:30 -- 13:00 &     KmSh       &       1920           & 2.3   &  4   & 1024    \\
RSY01    & 2011-10-17   & 16:10 -- 18:00 & HhWbYsTrOnMcEf &  266\,--\,8042       & 5.0   &  2   & 1024    \\
\hline
\end{tabular}
\end{table*}

\begin{table*}
\caption{The circular Gaussian model fitting results of the detected radio component in MAXI~J1836--194. }
\label{tab2}
\scriptsize
\setlength{\tabcolsep}{10pt}
\centering
\begin{tabular}{ccccccc}

\hline
Frequency & MJD       & TECOR  &  Right Ascension                        & Declination                       &   Flux Density & Angular Size\\
(GHz)     & (day)     &        &                                &                                   & (mJy)          & (mas)       \\
\hline
8.3       & 55844.450 & None   & $18^\mathrm{h}35^\mathrm{m}43\fs44439$ &  $-19\degr19\arcmin10\farcs4825$  &  $6.46\pm0.33$   & $\leq0.5$   \\
8.3       & 55844.450 & JPLG   & $18^\mathrm{h}35^\mathrm{m}43\fs44451$ &  $-19\degr19\arcmin10\farcs4869$  &  $6.64\pm0.32$   & $\leq0.5$   \\
8.3       & 55844.450 & CODG   & $18^\mathrm{h}35^\mathrm{m}43\fs44452$ &  $-19\degr19\arcmin10\farcs4868$  &  $7.23\pm0.30$   & $\leq0.5$   \\
\hline
5.0       & 55851.712 & None   & $18^\mathrm{h}35^\mathrm{m}43\fs44449$ &  $-19\degr19\arcmin10\farcs4917$  &  $4.09\pm0.56$   & $\leq0.4$   \\
5.0       & 55851.712 & JPLG   & $18^\mathrm{h}35^\mathrm{m}43\fs44451$ &  $-19\degr19\arcmin10\farcs4923$  &  $4.33\pm0.45$   & $\leq0.4$   \\
5.0       & 55851.712 & CODG   & $18^\mathrm{h}35^\mathrm{m}43\fs44455$ &  $-19\degr19\arcmin10\farcs4980$  &  $4.03\pm0.57$   & $\leq0.4$   \\
\hline
\hline
\end{tabular}
\end{table*}

\section{Observations and Data Reduction}
\label{sec2}

\subsection{CVN observations at 3~Gbps recording rate}
\label{sec2-1}

We conducted the first-epoch observations of MAXI~J1836--194 during a dual-frequency, wide-band test experiment with the Chinese VLBI Network (CVN) on 2011 October 10 (project code CHIN06; Table~\ref{tab1}). Three antennas participated: Shanghai/Sheshan 25m radio telescope, Urumqi/Nanshan 25m radio telescope, and Kunming 40m radio telescope \citep[a new VLBI station in Yunnan Province, see][]{hao10}. Each radio telescope observed only right-hand circular polarisation and used two back-ends: Mark4 at 2.3~GHz and CDAS \citep[Chinese VLBI Data Acquisition System System,][]{zha10} at 8.3~GHz. Both back-ends were configured to their maximum recording rate: 1024~Mbps ($16\times16$~MHz, 2-bit sampling) for Mark4, and 2048 Mbps ($16\times32$~MHz, 2-bit sampling) for CDAS.

We selected PMN~J1832--2039 as the phase-reference calibrator source. Its J2000 position according to the global astrometric solution rfc\_2010c\footnote{http://astrogeo.org/rfc} is: $\alpha = 18^\mathrm{h}32^\mathrm{m}11\fs046479$ ($1\sigma=0.2~\mathrm{mas}$), $\delta$ = $ -20\degr39\arcmin48\farcs20367$ ($1\sigma=0.4~\mathrm{mas}$). It lies $1.\!^\circ6$ away from the target, to the south-south-west. The duty cycle time was 3 minutes: 1 minute on the calibrator and 2 minutes on the target. We observed this pair of sources for four hours. The remaining two hours in the experiment were used to observe OQ208 to verify the performance of the CDAS back-end. The data were correlated with the EVN Software Correlator at JIVE (SFXC). The correlation parameters were 1 second integration time and 256 frequency points per subband. The Urumqi 2.3-GHz data were lost due to a recording failure.

\subsection{EVN observations in e-VLBI mode}
\label{sec2-2}
We performed the second-epoch observations with the EVN at 5~GHz on 2011 October 17 (EVN project code RSY01; see Table~\ref{tab1}). The participating stations were Effelsberg, the phased array Westerbork Synthesis Radio Telescope (WSRT), Onsala, Medicina, Hartebeesthoek, Yebes, and Torun.  The observing set-up was 1024~Mbps, using both circular polarizations. The phase-reference source remained PMN~J1832--2039 for these observations. Furthermore, we also observed two nearby sources: NVSS~J183542--191943 (0.5 arcminutes away from the target, 35.1~mJy at 1.4~GHz) and NVSS~J183522--191743 (5 arcminutes away, 28.7~mJy at 1.4~GHz) in the hope that these could be used as secondary phase-reference calibrators. The duty cycle time was 5 minutes: 1 minute spent on the main calibrator, 1.5 minute on the transient, and 1.5 minute on one of the NVSS sources (each alternating in successive phase-reference cycles). This epoch lasted for two hours. The data were streamed to JIVE and correlated in real-time (e-VLBI) using the EVN MkIV Data Processor with 1 second integration time and 64 frequency points per subband/polarisation.

\begin{figure*}
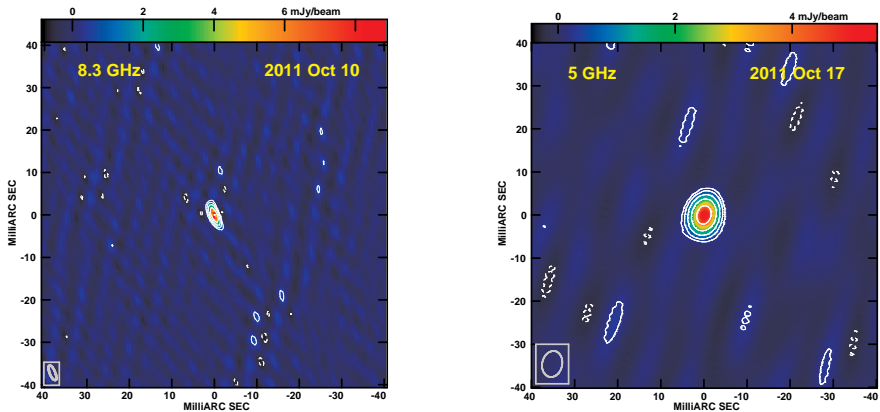

  \centering
  \includegraphics[width=0.3\textwidth]{fig1a.eps}
  \hspace{1cm}
  \includegraphics[width=0.3\textwidth]{fig1b.eps} \\
  \caption{VLBI total intensity images of the Galactic black hole binary candidate MAXI~J1836--194.
Both images were produced with natural weighting. The synthesised beams are plotted in the bottom-left
corner (left image: $3.7\times1.3$~mas at the position angle $21.8$ degree, right image:
$6.1\times4.6$~mas at the position angle $-12.9$ degree). The contours start from $3\sigma$ off-source
noise level (0.54 mJy\,beam$^{-1}$ at 8.3 GHz, 0.24 mJy\,beam$^{-1}$ at 5 GHz) and increase by a
factor of 2. } \label{fig1}
\end{figure*}

\subsection{Data reduction}
\label{sec2-3}

We calibrated the data using the NRAO Astronomical Image Processing System \citep[AIPS, version 31DEC11,][]{gre03}, using the following procedure.
(1) We removed the ionospheric delay using total electron content (TEC) measurements from GPS monitoring. (2) We calibrated the correlation amplitudes based on the measured system temperatures and antenna gain curves.  We used nominal system equivalent flux densities (SEFD) for Kunming and Shanghai. (3) We corrected for the antenna parallactic angle. (4) We corrected the position of the Kunming telescope with the AIPS task CLCOR to account for improvement in its position estimate that became available after the correlation ($\Delta\,X=0.157$~m, $\Delta\,Y=0.486$~m, $\Delta\,Z=0.458$~m). (5) We then fringe-fitted the PMN~J1832$-$2039 data, and corrected the antenna bandpasses based on the calibrator as well. (6) We imaged PMN~J1832--2039 in Difmap \citep{she94}, after all the calibration solutions were applied. (7) We removed the phase contribution arising from the source structure through a further step of global fringe fitting using the calibrator image in AIPS. (8) We transferred the phase solutions to the target sources via linear interpolation. (9) We self-calibrated the calibrator and applied its amplitude and phase solutions to the target source.  Note that the amplitude self-calibration was not possible in the 8.3 GHz experiment because of the small number ($<4$) of the stations. We then averaged the data in frequency over all subbands. (10) We identified the transient source in the dirty map and measured its position by fitting a point source model to the visibility in Difmap. (11) After the model position was fixed, we self-calibrated the data using a 2-minute solution interval to improve the noise level and look for possible other components. We kept Hartebeesthoek (long baselines) and Onsala (less sensitive) phases fixed in this process.

We also reduced the 5-GHz WSRT synthesis array data in AIPS, using 3C~286 to calibrate the flux densities. We self-calibrated PMN~J1832--2039 (an ideal point source for the WSRT) in both phase and amplitude, and applied the solutions to the target. Finally, we averaged the data in each subband and imaged the source in Difmap.

\section{VLBI imaging results}
\label{sec3}

The VLBI imaging results are displayed in Figure~\ref{fig1}. MAXI~J1836--194 is unresolved at both frequencies; the visibility data can be fitted with a point source model very well. The model-fitting results are listed in Table~\ref{tab2}. The phase-reference calibrator image is shown in Figure~\ref{fig2}. The calibrator source had a bright, one-sided jet. The eastern component has the most compact structure with similar peak brightness ($\sim$0.38~Jy\,beam$^{-1}$) at the two frequencies. Thus, we assume this to be the radio core. We measured the source position with respect to the compact core of the calibrator at each frequency. Note that this new position of the target source is different from the one we reported earlier \citep{xu11}: in that telegram, the reference origin was the central bright jet component rather than the radio core. In order to estimate the upper limit of the angular size of MAXI~J1836--194, we also fitted the visibility data with a circular Gaussian model. The last column in Table~\ref{tab2} presents the estimates of the angular size. Both observations give a consistent upper limit of 0.5~mas.

We also searched for extended emission in the VLBI image by increasing the weights of short-baseline data. There was no extended emission found. Based on the WSRT observations, the total flux density of the transient at 5~GHz was $5.7\pm0.2$~mJy. Therefore we recovered 95\% of the total flux density in the simultaneous VLBI observations. If there is extended emission, its contribution to the total flux density is quite low, $\leq5$\%.

There was only one baseline (Kunming~--~Shanghai) at 2.3 GHz. The baseline gave a total flux density $5\pm1$~mJy. Because phase self-calibration was not possible, there may have been a significant loss of phase coherence at this frequency (note the very low declination). Assuming $\sim$30\% loss, similar to what we found after applying the phase self-calibration at 8.3~GHz, the source would have a flux density close to that at 8.3~GHz. Thus, the source did not have a spectral index significantly different from 0 on the date of these dual-frequency observations.

The two nearby NVSS sources that we included in the 5~GHz observations as candidate secondary calibrators within $5\arcmin$ of MAXI\,J1836--194 were not detected at the threshold of $5\sigma\sim1.3$~mJy.  Therefore, they could not be used to estimate spatial gradients of propagation effects to the phase referencing in the vicinity of the target.

\section{Discussion}
\label{sec4}

\subsection{MAXI J1836--194 position stability}
 \label{sec4-1}

\begin{figure}
  \centering
  \includegraphics[width=0.3\textwidth]{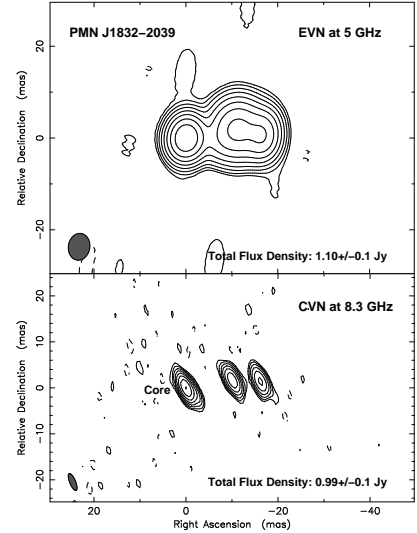} \\
  \caption{VLBI total intensity images of the phase-referencing source PMN~J1832--2039. Both images were produced with natural weighting. The synthesised beams are plotted in the bottom-left
corner. The contours start from $3\sigma$ off-source noise level (0.5 mJy\,beam$^{-1}$ at 5 GHz, 1.0 mJy\,beam$^{-1}$ at 8.3 GHz) and increase by a factor of 2. } \label{fig2}
\end{figure}

After careful removal of source-structure effects in the phase-referencing (c.f. Sect.~\ref{sec2-3}, steps 6--8), there is still a notable position difference ($\Delta\alpha$=0--1.4~mas and $\Delta\delta$=5.4--11.2~mas) between 5 and 8.3~GHz. There may well be a frequency-dependent core shift in the reference source J1832--2039 that could not calibrate absolutely with available observations, but this would lie along the E-W jet axis (see Fig.~\ref{fig2}) and generally be on the order of 0.1~mas between 5--8.3~GHz \citep{kov08}. Such a core shift would not explain the apparent position difference which is predominantly in declination. Other critical characteristics of these observations include the short duration and the very southerly target (and an even more southerly reference source), leading to persistently low elevations from European VLBI antennas.  At these low elevations, even the $1.\!^\circ6$ separation between the target and reference sources is larger than ideal.  Thus the first suspects would be imperfectly calibrated propagation effects in the phase-referencing.

These observations from October 2011 took place at a time of relatively elevated solar activity coming out of the prolonged quiet period, which could lead to a more active ionosphere. Quantitatively, the daily solar 10.7cm flux density (e.g., NOAA National Geophysical Data Centre), averaged over a 27-day solar rotation period was about 140 [$10^4$ Jy], a value not seen since late 2003. The EVN MkIV Data Processor applies no {\it a priori} ionospheric delays; rather it is standard to apply IONEX maps from the International GNSS Service (IGS) \citep{igs09, rcw99} in AIPS data reduction, as described in Section~\ref{sec2-3}.  There are a few IGS analysis centres who process observations of GPS satellites from 100--150 ground stations to produce these IONEX maps of vertical column density of electrons, gridded spatially at ($\Delta\Phi=2\fdg5$, $\Delta\lambda=5\fdg0$) and temporally at 2-hour intervals.  The AIPS task TECOR computes phase corrections for the lines of sight from the VLBI antennas to the sources throughout the observations by interpolating within these IONEX maps. Table~\ref{tab2} shows the effect of treating the ionospheric delays in three ways: making no correction (``None"), and then using the IONEX maps from the analysis centres at JPL (``JPLG") and the University of Bern (``COGD"). Using IONEX maps from these different Analysis Centres leads to changes of up to 6 mas in the declination of the target, which is on the order of the observed position difference.

The correlator model computed dry and wet tropospheric delays from the Saastamoinen zenith-delay formulation \citep[e.g.,][]{jld85}, parameterized by surface pressure, temperature, and relative humidity, and the Neill mapping function \citep[NMF,][]{aen96} to convert from zenith to slant delays along the lines of sight.  At low elevations, mismodelling of the zenith delays can become amplified ($\sec{z}$ for the RSY01 observation average 2--5 for the various European VLBI antennas). \citet{bru05} developed a method to determine improved zenith-delay estimates via scheduling geodetic-VLBI-like blocks into the observations, but the short duration of the present experiments precluded use of this tactic. To characterise the range of position shifts that may be induced by mismodelled tropospheric delays, we computed zenith delays in two ways: using the surface-weather values for antennas that had such information in their log files and using Calc-10 \citep{gor06} defaults. We applied the NMF to both, and computed differential tropospheric delays.  The median of these was up to $\sim$50~ps, corresponding to a phase offset $\sim$100$^\circ$ at 5~GHz. Through adjusting the position of the target and running phase self-calibration, we found that similar phase change results from a position shift of $\sim$5~mas.

Thus the variation in the position of the target estimated from the CHIN06 and RSY01 observations is consistent with effects of ionospheric/tropospheric propagation in the phase referencing. Similar effects were also noticed in other low-declination sources, such as H\,1743--322 \citep{mil12}.

Nonetheless, we can also briefly consider physical causes of a position shift associated with the target itself over the week between the two observations. The transient produced a ``failed" outburst \citep{fer12}, never transiting to the high/soft X-ray state during its evolution (low/hard, hard intermediate, and back to low/hard X-ray states). Both of our observations were carried in the canonical low/hard state, where major ejection events are unlikely \citep{cor04,fen04,fen09}. Moreover, our target source was quite compact at both epochs with brightness temperatures $\sim$10$^8$~K, showing no sign of extension between the two epochs. That the observed spectrum between 2.3 and 8.3 GHz is consistent with being optically thick indicates that we detected a compact, synchrotron self-absorbed component of the system---its radio core. This is in agreement with the fast variability observed by the RATAN-600 \citep{tru11} during the hard state at an earlier epoch. Thus the position shift is unlikely to be related to moving ejecta.

The position measured at 8.3 GHz is likely more reliable on several grounds: the beam pattern of the 8.3~GHz data was clearly seen in the dirty map, the underlying CVN observations had twice the on-source time and elevations that were generally 10$\degr$ higher, and the higher frequency affords 40\% less ionospheric-phase contribution for comparable TEC levels.

\subsection{VLBI detections as the compact-jet region}
\label{sec4-2}

The arguments presented above convince us that the point-like detections in Figure~\ref{fig1} correspond to the same object, i.e., the radio core of MAXI~J1836--194. The existence of a synchrotron jet in the vicinity of the stellar-mass compact object in MAXI~J1836--194 is further supported by the optically thick spectrum in the radio \citep{tru11}, and the optically thin power-law spectrum in the mid-infrared \citep{rus11}. Similar compact jets have been directly resolved in the hard state in black hole X-ray binaries, e.g., Cygnus~X--1 \citep{sti01} and GRS~1915$+$105 \citep{dha00}. In the present case of MAXI~J1836--194, the compact jet remains unresolved. We can constrain its angular size to be $\leq$0.5~mas, which corresponds to an AU-scale linear size for a Galactic source. Because MAXI~J1836--194 lies outside of the Galactic plane ($b=-5\fdg4$), this constraint should not be affected by scatter-broadening arising from the Galactic medium along the light of sight, which could potentially smear out the intrinsic source structure at low frequencies.
We compare the apparent size constraint with the known, resolved compact jets below.

Within the framework of the jet model of \citet{fal99} with minimal number of free parameters, the characteristic size $\theta_\mathrm{jet}$ of the compact jet significantly depends on the mechanical power of the jet: $\theta_\mathrm{jet} \propto Q_\mathrm{jet}^{0.6\chi_1}$, where $\chi_1$ is a parameter depending on the inclination angle of the jet, with a value generally around 1, and the jet power $Q_\mathrm{jet}$ is on the order of the X-ray luminosity of the coupled disk-jet system considered by \citet{fal99}. MAXI~J1836$-$194 had an X-ray luminosity $\sim10^{36}$~erg\,s$^{-1}$ between 0.3~--~10~keV \citep[corrected for absorption,][]{ken11} assuming a distance of 8.5~kpc, three orders of magnitude lower than that ($\sim10^{39}$~erg\,s$^{-1}$) of GRS~1915$+$105. Assuming that the parameters not determined from these observations (distance, mass and inclination angle) are the same as GRS~1915$+$105 \citep{dha00}, MAXI~J1836--194 would have a characteristic size at least an order of magnitude smaller than that of GRS~1915$+$105 ($\sim$0.7~mas at 8.3~GHz). This result is consistent with our upper limit of the source size. We therefore conclude that we have detected the unresolved, synchrotron self-absorbed jet base in MAXI~J1836--194, in accordance with the expectations for its hard/low X-ray spectral state.

\section{Conclusions}
\label{sec5}
We have presented VLBI observations of the recently detected Galactic variable X-ray source MAXI~1836--194 with the EVN at 5~GHz and the CVN at 2.3/8.4 in October 2011. Despite the unfavourable celestial position for the telescopes located in the Northern hemisphere, this low-declination Southern object has been firmly detected at all three frequencies as an unresolved radio source. The VLBI data allow us to put a firm upper limit on the angular size of the source as 0.5~mas, which corresponds to an AU-scale linear size for a Galactic object and is as expected within the framework of compact synchrotron jet models.

\section*{Acknowledgments}
\label{ack}
\footnotesize
We thank the EVN Programme Committee Chair for approving our short e-VLBI observations RSY01, and for permitting us to use the CVN CDAS test experiment CHIN06 to observe MAXI~1836$-$194. e-VLBI research infrastructure in Europe is supported by the European Union's Seventh Framework Programme (FP7/2007-2013) under grant agreement RI-261525 NEXPReS. We acknowledge support from the Royal Dutch Academy of Sciences (KNAW) and the Chinese Academy of Sciences (CAS) via their joint (project no. 10CDP005). The EVN is a joint facility of European, Chinese, South African and other radio astronomy institutes funded by their national research councils. The WSRT is operated by ASTRON with support from the Netherlands Foundation for Scientific Research.

\bsp
\label{lastpage}
\end{document}